\documentclass[twocolumn,pre,superscriptaddress,amsmath,amssymb]{revtex4}
\usepackage{graphicx}  
\usepackage{color}

\begin{document}
\title{Active matter: quantifying the departure from equilibrium}
\date{\today}

\author{Elijah Flenner}
\affiliation{Department of Chemistry, Colorado State University, 
Fort Collins, Colorado 80523, USA}
\author{Grzegorz Szamel}
\affiliation{Department of Chemistry, Colorado State University, 
Fort Collins, Colorado 80523, USA}

\begin{abstract}
Active matter systems are driven out of equilibrium at the level of individual
constituents. One widely studied class are systems of athermal particles that
move under the combined influence of interparticle interactions and self-propulsions,
with the latter evolving according to the Ornstein-Uhlenbeck stochastic process. 
Intuitively, these so-called active Ornstein-Uhlenbeck particles (AOUPs) systems are 
farther from equilibrium for longer self-propulsion persistence times. Quantitatively,
this is confirmed by the increasing equal-time velocity correlations 
(which are trivial in equilibrium) and by the increasing violation of the Einstein 
relation between the self-diffusion and mobility coefficients. In contrast, the entropy
production rate, calculated from the ratio of the probabilities of the position space
trajectory and its time-reversed counterpart, has a non-monotonic dependence
on the persistence time. Thus, it does not properly quantify the departure of
AOUPs systems from equilibrium.  
\end{abstract}

\maketitle

\section{Introduction}
The focus of our study are model active matter systems consisting of 
athermal self-propelled particles, which move due to interparticle 
interactions and self-propulsions, with the self-propulsions evolving independently of 
the positions of the particles \cite{Bechingerrev,Marchettirev2}. To fully define these 
systems one needs to specify the properties of the self-propulsions.
Two popular choices are active Brownian particles (ABPs) \cite{tenHagen,FilyMarchetti}
which are endowed with self-propulsions
of a constant magnitude and evolving via rotational diffusion, and active 
Ornstein-Uhlenbeck particles (AOUPs) \cite{Szamel2014,Maggi2015,Fodor2016}
for which the self-propulsions evolve according
to the Ornstein-Uhlenbeck stochastic process. 
Since our fundamental understanding of non-equilibrium systems is not well
developed, one often used approach to study systems of self-propelled
particles is to approximate them by appropriately chosen equilibrium systems
\cite{Maggi2015,Das2014,FarageKB}. 
Intuitively, whether such an approach is sensible depends on how non-equilibrium the
former systems are \cite{noneqeq}. 
Thus, the recurring question in the recent literature is how 
to quantify the departure of the systems of self-propelled particles 
from equilibrium. To paraphrase a recent article \cite{Li2019}, 
we want to replace a binary in-out 
of equilibrium classification by a more quantitative one.

One possibility is to generalize the stochastic thermodynamics
approach and to evaluate the entropy production defined through
a ratio of the probability of the forward trajectory and that of its 
time-reversed version. For systems of AOUPs this approach was first proposed 
by Fodor \textit{et al.} \cite{Fodor2016} and then elaborated on by 
Puglisi, Marconi, Maggi and collaborators \cite{Marconi2017,Caprini2019}.
They defined the entropy production in terms of the ratio of the probabilities
of the forward and reversed trajectories in the position space 
and derived a compact
expression which makes a numerical evaluation of the entropy production straightforward.

There have been other attempts to define the entropy production.
Mandal \textit{et al.} \cite{Mandal2017} defined the entropy 
production in terms of the ratio of the probabilities
of the forward trajectory and a trajectory following 
time-reversed evolution. Dabelow \textit{et al.} \cite{Dabelow2019} argued that 
the relation between the entropy production and the trajectory probability ratio involves
an additional quantity originating from the ``mutual information'' 
between the trajectory and the environment. Shankar and Marchetti \cite{Shankar2018}
proposed calculating the entropy production from the ratio of the probabilities
of the forward and reversed trajectories in the enlarged phase space consisting
of the particle's position and self-propulsion. While they considered only 
a single free self-propelled particle, their approach was 
generalized by one of us to a single AOUP in an external potential \cite{Szamel2019}.
Additionally, Pietzonka and Seifert \cite{Pietzonka2018} 
argued that the most fundamental consideration
of entropy production should also include the contribution from 
physico-chemical processes that give rise to the self-propulsion. 
 
Here we present results of a simulational investigation of the entropy production
according to Fodor \textit{et al.} for systems of interacting AOUPs. The most
intuitive control parameter tuning the departure of these systems from equilibrium
is the persistence time of the self-propulsion. We present quantitative numerical
results supporting this expectation. Then, we show that the expression 
for the entropy production derived in Ref. \cite{Fodor2016} has a non-monotonic
dependence on the self-propulsion persistence time. Thus, it
is not a good measure of the departure of AOUP systems from equilibrium.
Numerical evaluation of the alternative proposals to define the entropy production 
\cite{Mandal2017,Dabelow2019,Shankar2018,Szamel2019} is left for a future study.

\section{Simulations} 
We simulated interacting, athermal AOUPs
\cite{Szamel2014,Maggi2015,Fodor2016}, moving
in a viscous medium, without inertia, under the combined influence of the interparticle
forces and self-propulsions, with the latter evolving according to 
the Ornstein-Uhlenbeck stochastic process.
The equations of motions read
\begin{eqnarray}
\label{eq:motion1}
\dot{\mathbf{r}}_i &=& \xi_0^{-1}[\mathbf{F}_i + \mathbf{f}_i], \\
\tau_p \dot{\mathbf{f}}_i &=& \mathbf{f}_i + \boldsymbol{\eta}_i. 
\label{eq:motion2}
\end{eqnarray}
In Eq.~\eqref{eq:motion1} $\mathbf{r}_i$ is the position of particle $i$, 
$\xi_0$ is the friction coefficient of an isolated particle, $\mathbf{F}_i$ is 
the interparticle force, and $\mathbf{f}_i$ is the self-propulsion. 
In Eq.~\eqref{eq:motion2} $\tau_p$ is the persistence time
of the self-propulsion and $\boldsymbol{\eta}_i$ is the internal Gaussian noise 
with zero mean and variance 
$\left< \boldsymbol{\eta}_i(t) \boldsymbol{\eta}_j(t^\prime) \right>_{\mathrm{noise}} 
= 2 \xi_0 k_B T_a \mathbf{I} \delta_{ij} \delta(t-t^\prime)$,
where $\left< \ldots \right>_\mathrm{noise}$ denotes averaging over the noise 
distribution, $T_a$ is the ``active'' temperature, and $\mathbf{I}$ is the unit tensor. 
We choose a system of units such that $\xi_0=1$ and $k_B=1$. 
We emphasize that $T_a$
characterizes the strength of the self-propulsion; it is called the active temperature
because it determines the long-time diffusion coefficient of a single free AOUP,
$D_0=k_B T_a/\xi_0\equiv T_a$.  

We studied a 50:50 binary mixture of $N=1000$ particles interacting via the smoothed 
Weeks-Chandler-Andersen truncation of the Lennard-Jones potential,
$V_{\alpha \beta}(r) = 4 \epsilon 
\left[ \left(\frac{\sigma_{\alpha \beta}}{r}\right)^{12} 
- \left(\frac{\sigma_{\alpha \beta}}{r}\right)^6\right]+V_{\alpha \beta}^\text{cut}(r)$,
where $\alpha, \beta$ denote the particle species $A$ or $B$, $\epsilon = 1$, 
$\sigma_{AA} = 1.4$, $\sigma_{AB} = 1.2$, $\sigma_{BB} = 1.0$, and 
$V_{\alpha \beta}^\text{cut}(r)=c_0+c_2 \left(r/\sigma_{\alpha \beta}\right)^2+
+c_4 \left(r/\sigma_{\alpha \beta}\right)^4+c_6 \left(r/\sigma_{\alpha \beta}\right)^6$.
The potential is truncated and shifted at 
$\varsigma_{\alpha \beta} = 2^{1/6} \sigma_{\alpha \beta}$ and the parameters
$c_0$, $c_2$, $c_4$, and $c_6$ are chosen
such that the potential and its first three derivatives are continuous at the cutoff.
The resulting inter-particle force 
$\mathbf{F}_i = - \sum_{j \ne i} \partial_{\mathbf{r}_i} 
V_{\alpha \beta}(r_{ij})$ is purely repulsive. 
All the quantities presented in this work except
for the velocity correlations, Eq.~(\ref{ompar}), pertain to all, \textit{i.e.} 
large and small, 
particles. The velocity correlations were calculated for the large particles only;
the correlations for the small particles are qualitatively the same.

Our control parameters were the active temperature $T_a$, 
the packing fraction $\phi = \pi N[\varsigma_{AA}^3 + \varsigma_{BB}^3]/(12 V)$ 
and the persistence time $\tau_p$.
We performed simulations along two lines in this three-dimensional space,
specified by 
[$T_a = 1.0$, $\phi = 0.64$] and [$T_a = 0.01$, $\phi = 0.58$]. As a shortcut,
we refer to these two lines as two ``state points'', in spite of the fact that
the full specification of the state point requires also $\tau_p$.  
When the persistence time goes to zero at a fixed active temperature our system
becomes equivalent to a Brownian system at temperature $T = T_a$. 
We chose the two state points in such a way that we could 
observe qualitatively different changes of the single-particle dynamics
with increasing persistence time \cite{BFS2017}. 

\section{Equal time velocity correlations}
Intuitively, by increasing persistence time we displace an AOUP system farther from
equilibrium. To give some quantitative support to this statement we
investigated equal time correlations of the velocities of the active particles. 
These correlations are trivial for equilibrium thermal systems. We
note that non-trivial equal time velocity correlations were observed in an 
experimental study of active cellular motion \cite{Garcia2015} and in a simulational
investigation of the dense phase in an active system undergoing 
mobility-induced phase separation \cite{Caprini2020}. 

In our earlier investigations of glassy dynamics in interacting AOUP systems 
\cite{BFS2017,SFB2015,Szamel2016} we found that the 
equal time velocity correlation function defined below determines the short-time
dynamics of the active particles and also appears in an approximate mode-coupling-like
theory of the long time dynamics, 
\begin{equation}\label{ompar}
\omega_{\parallel}(q) = \frac{1}{N\xi_0^2} 
\left< \left| \sum_i
\hat{\mathbf{q}} \cdot \left(\mathbf{f}_i+\mathbf{F}_i\right)
e^{-i\mathbf{q}\cdot\mathbf{r}_i}\right|^2\right>.
\end{equation}
Here 
$\hat{\mathbf{q}} = \mathbf{q} / | \mathbf{q} |$ and 
$\xi_0^{-1}\left(\mathbf{f}_i+\mathbf{F}_i\right)$
is the instantaneous velocity of particle $i$, see Eq.~(\ref{eq:motion1}).

\begin{figure}
\includegraphics[width=0.8\columnwidth]{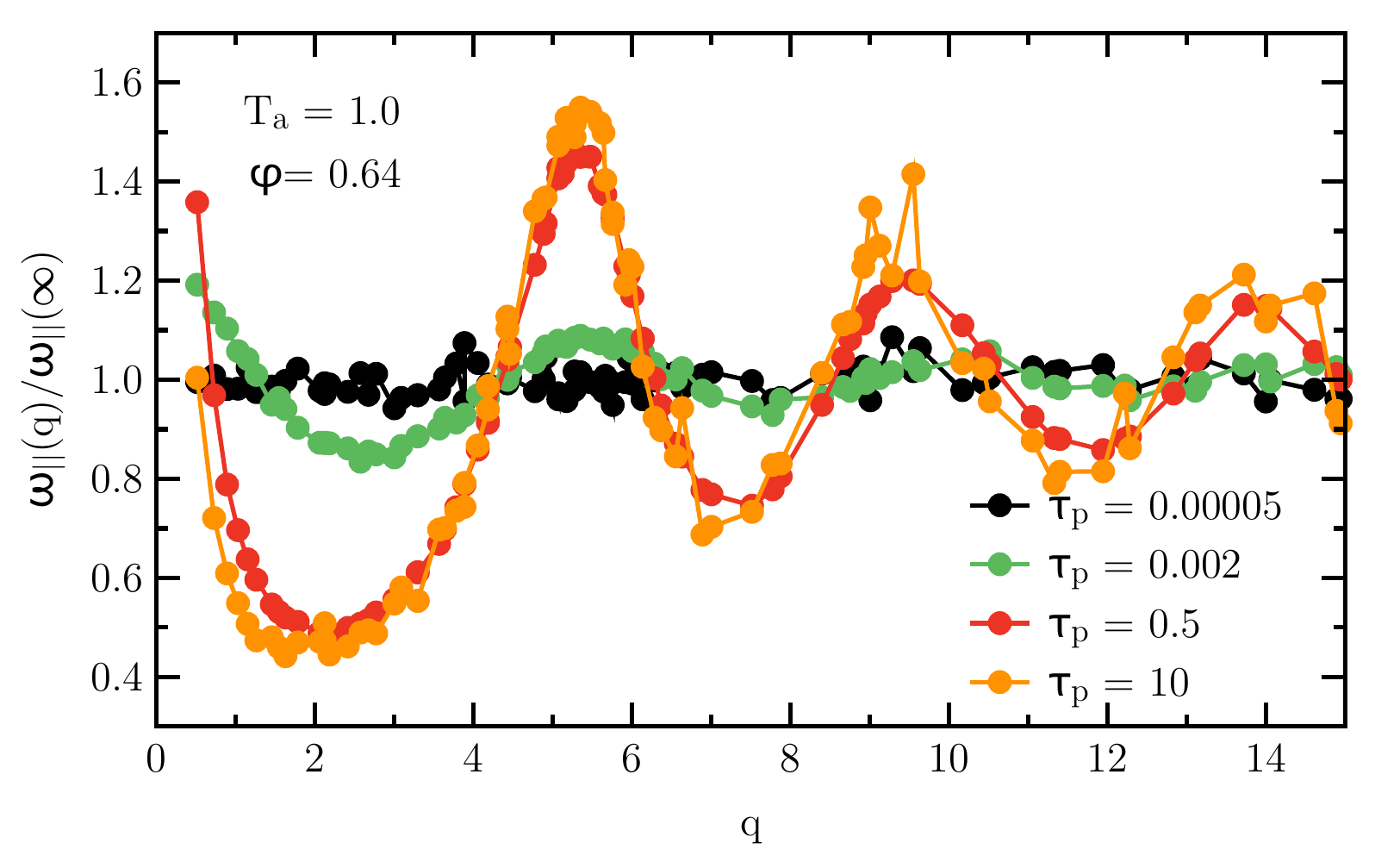}\\
\includegraphics[width=0.8\columnwidth]{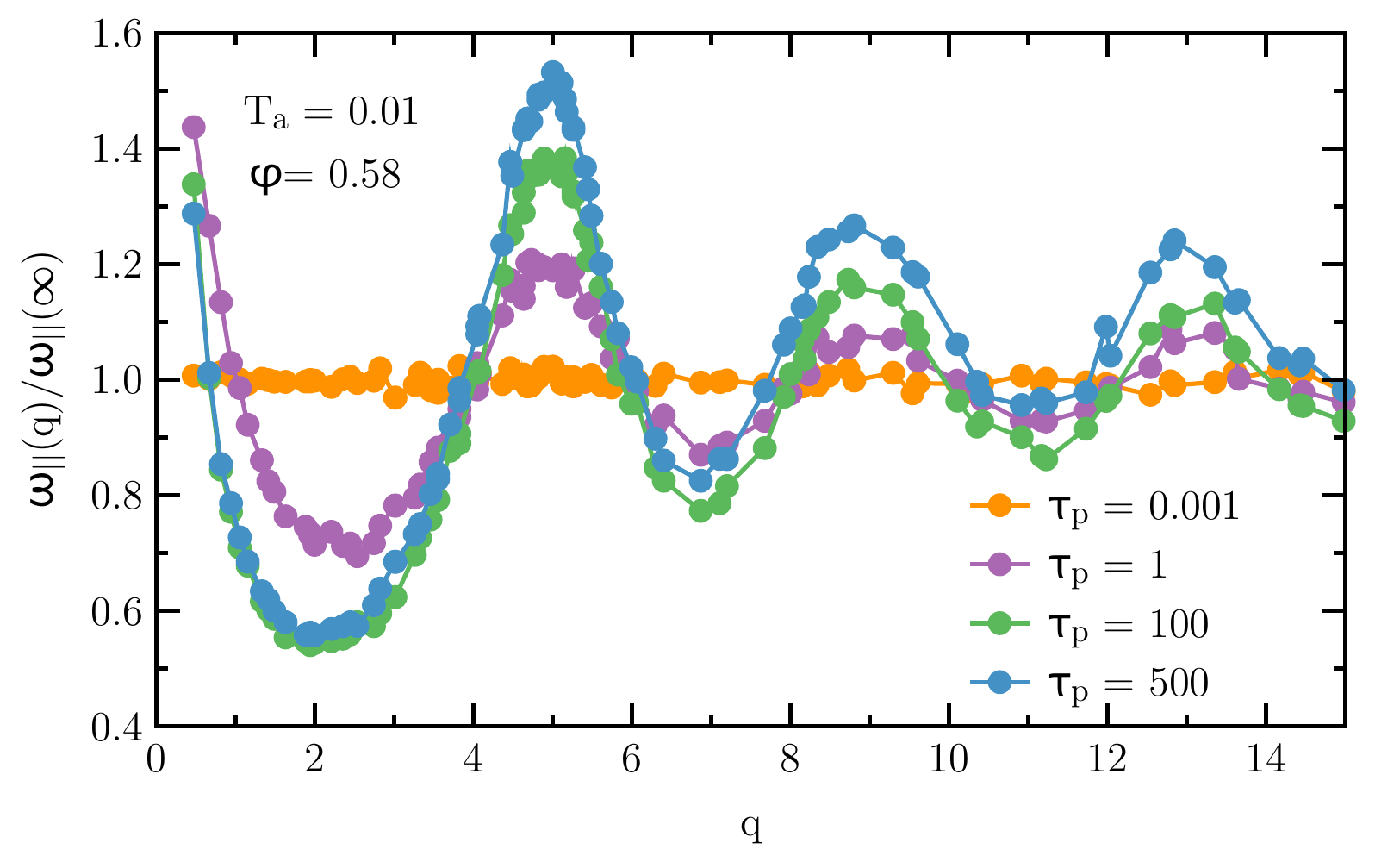}
\caption{\label{omega} The wavevector dependence of the equal-time nonequilibrium 
velocity correlation function $\omega_{\parallel}(q)$ normalized by its large wavevector 
limit, $\omega_{\parallel}(\infty)$, for (a) $T_a = 1.0$ and $\phi=0.64$ and 
(b) $T_a = 0.01$ and $\phi = 0.76$, 
and different self-propulsion persistence times.
Non-trivial character of the velocity correlations monotonically increases with
increasing persistence time.}   
\end{figure}

In Fig. \ref{omega} we show that, while velocity correlations as characterized by 
$\omega_{\parallel}(q)$ become trivial (featureless) 
in the limit of vanishing persistence time of the self-propulsions, 
they monotonically increase with the persistence time.

\section{Effective temperature based on the Einstein relation} 
The validity/violation of fluctuation-dissipation relations is a sensitive
signature for a system to be in/out of equilibrium \cite{Cugliandolorev}. 
To further verify that with
increasing persistence time a system of interacting AOPUs is progressively 
displaced away from equilibrium we test the validity of the simplest
fluctuation-dissipation relation between the self-diffusion and mobility 
coefficients and we compare the effective temperature defined as the ratio of these
coefficients to the active temperature.

\begin{figure}
\includegraphics[width=0.8\columnwidth]{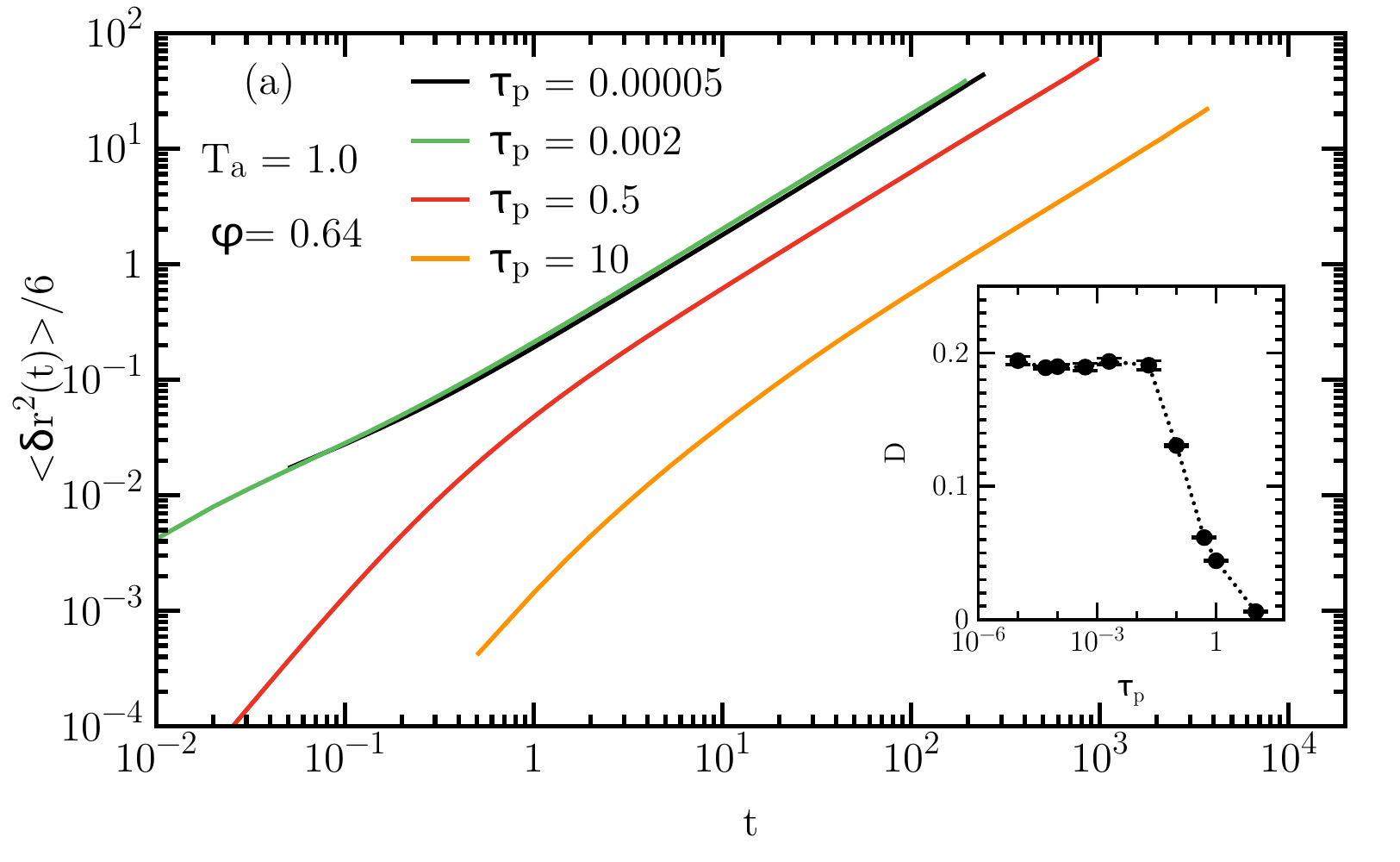}\\
\includegraphics[width=0.8\columnwidth]{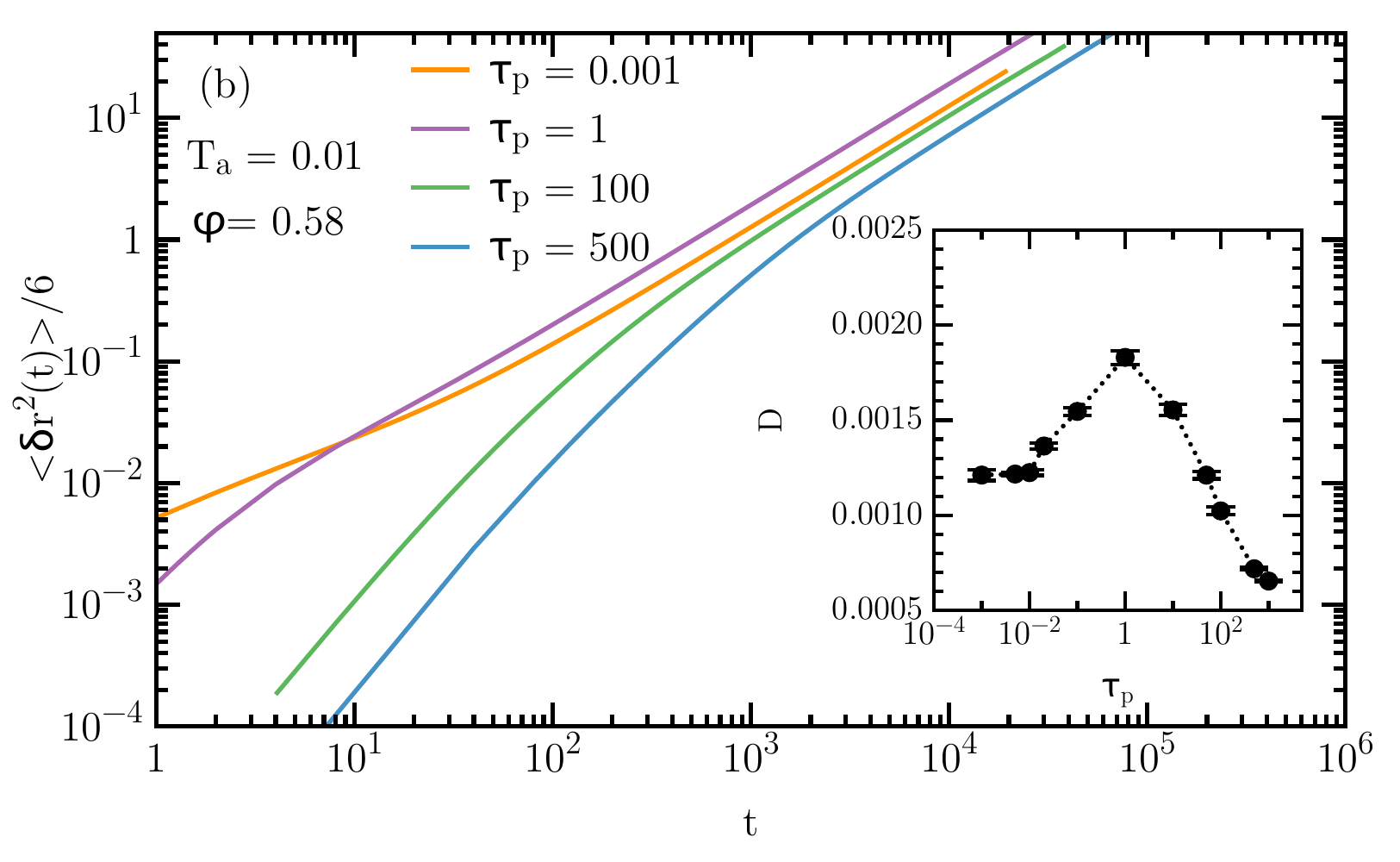}
\caption{\label{msd} The persistence time dependence of the mean-square displacement 
$\left<\delta r^2(t)\right>$ for (a) $T_a = 1.0$ and $\phi=0.64$ and 
(b) $T_a = 0.01$ and $\phi = 0.76$. The insets show the persistence time
dependence of the self-diffusion coefficient at each state point. With increasing
persistence time, the steady-state dynamics of the system studied at $T_a = 1.0$ 
monotonically slow down. However, with increasing persistence time
the system studied at $T_a = 0.01$ the steady-state dynamics initially speed up and 
then begin to slow down for the largest persistence
times studied.}
\end{figure}  

In Fig. \ref{msd} we show the persistence time dependence of the mean-square 
displacement, $\left<\delta r^2(t)\right> = 
N^{-1} \left< \sum_i \left(\mathbf{r}_i(t) - \mathbf{r}_i(0)\right)^2\right>$,
and of the self-diffusion coefficient, 
$D=\lim_{t\to\infty} \left<\delta r^2(t)\right>/(6t)$. 
We observe that at both state points $D$ is initially almost persistence 
time-independent (there is some slight non-monotonic dependence of $D$ on $\tau_p$, 
analogous to that reported earlier \cite{SFB2015}). Then, $D$ starts to decrease
rapidly with $\tau_p$ for the system at $T_a=1.0$. 
In contrast, for the system at $T_a=0.01$ $D$ starts to increase with increasing 
$\tau_p$. This behavior was previously observed in a similar system \cite{BFS2017}.
However, with increasing persistence time further than in this earlier study, 
the self-diffusion coefficient of the system at $T_a=0.01$ goes through a maximum and
starts decreasing. 

We define the time-dependent mobility function as follows: at $t=0$ a weak constant 
force $\lambda \mathbf{e}$ is applied to one (tagged, $t$) particle. 
Here $\lambda$ measures
the magnitude of the force and $\mathbf{e}$ is a unit vector. Under the influence of
this force the average position of the tagged particle will change systematically,
\begin{equation}\label{mobt}
\left<\mathbf{r}_t(t)-\mathbf{r}_t(0)\right> = \lambda\chi(t)\mathbf{e} + o(\lambda),
\end{equation}
where $\chi(t)$ is the time dependent mobility. We define the mobility
coefficient through the long-time limit of the time-dependent mobility function,
$\mu = \lim_{t\to\infty} \chi(t)/t$. We note that for an equilibrium system
the Einstein relation holds and $\left<\delta r^2(t)\right>/6 = T \chi(t)$, and 
$D=T\mu$, where $T$ is the system's temperature.

To calculate the time-dependent mobility of our active matter system 
we used the procedure proposed earlier
\cite{Szamel2017}, which allows one to evaluate a linear response function 
of an AOUP system using trajectories
generated without any external force. We note that, in general, 
longer trajectories are needed to accurately evaluate mobility function $\chi(t)$ than 
mean-square displacement $\left< \delta r^2(t) \right>$.

\begin{figure}
\includegraphics[width=0.8\columnwidth]{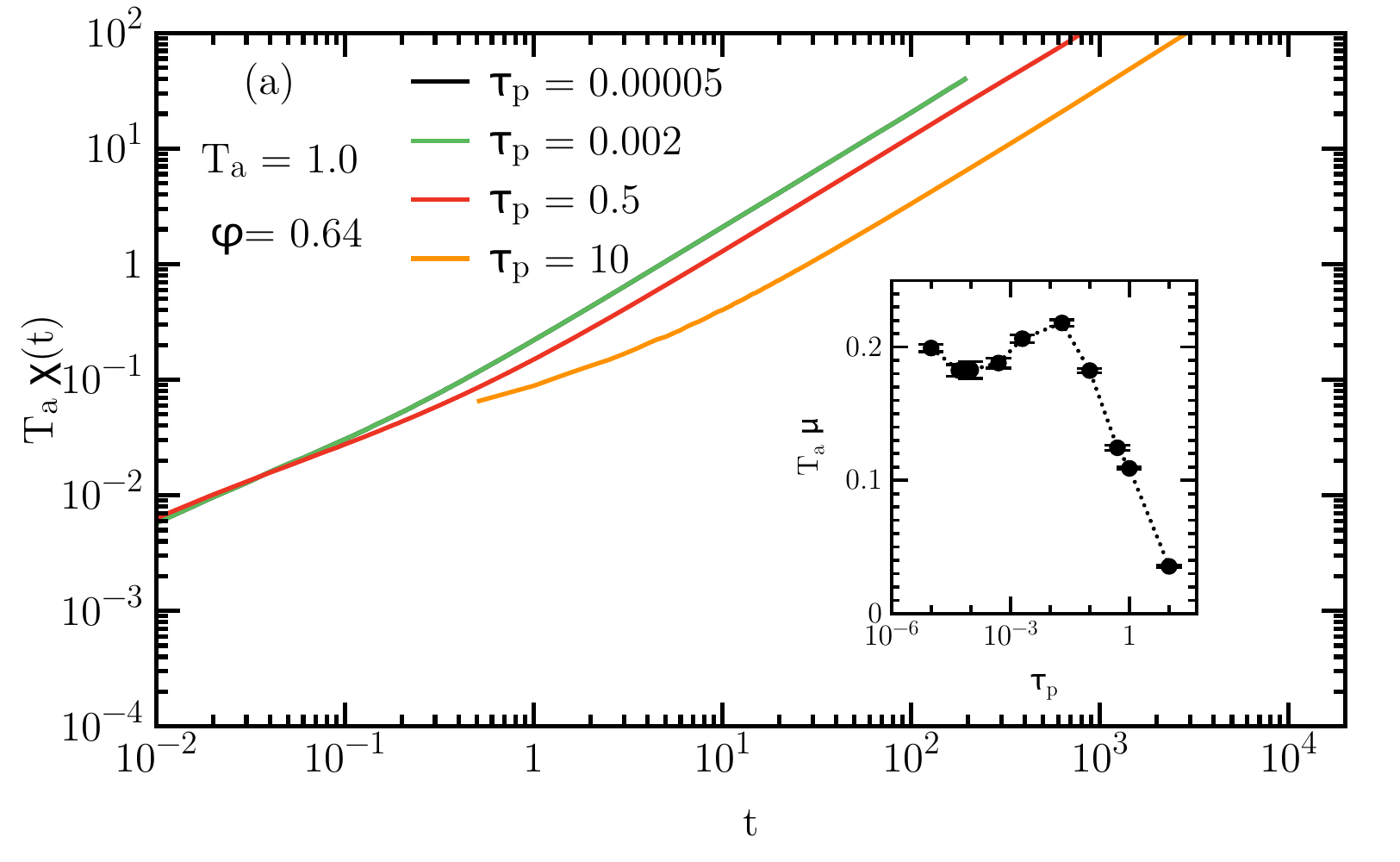}\\
\includegraphics[width=0.8\columnwidth]{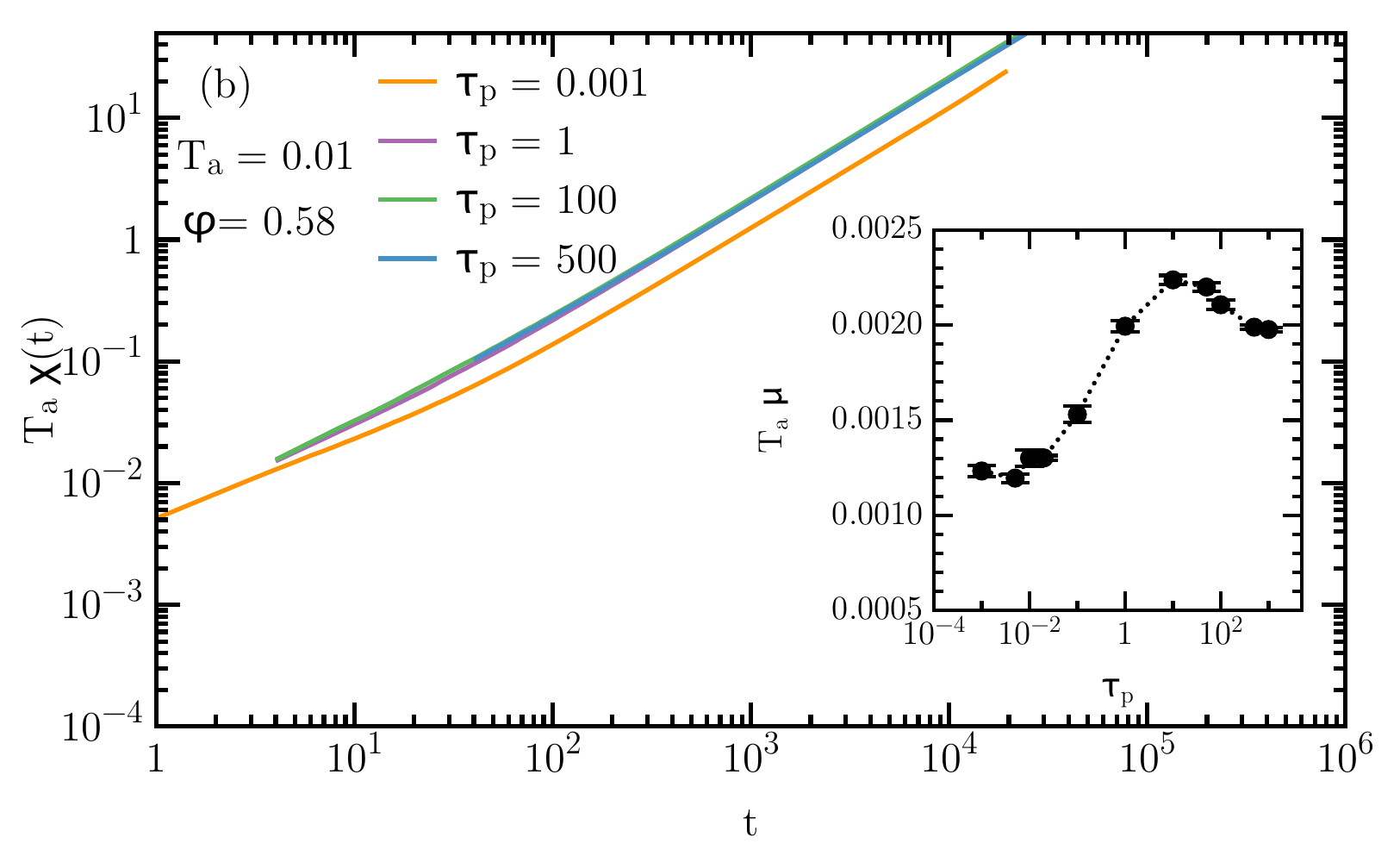}
\caption{\label{mobility} The persistence time dependence of the time-dependent
mobility function, $\chi(t)$,  for (a) $T_a = 1.0$ and $\phi=0.64$ and 
(b) $T_a = 0.01$ and $\phi = 0.76$. The insets show the persistence time
dependence of the mobility coefficient at each state point. With increasing
persistence time, the mobility of the system studied at $T_a = 1.0$ 
monotonically slows down. In contrast, with increasing persistence time
the system studied at $T_a = 0.01$ the mobility initially speeds up and 
then begins to slow down for the largest persistence
times studied. }
\end{figure}

In Fig. \ref{mobility} we show the persistence time dependence of the 
time-dependent mobility function and of the mobility coefficient at both state 
points. Comparing the main panels of Figs. \ref{msd} and \ref{mobility} we
see that the time-dependence of $\left< \delta r^2(t) \right>$ and $\chi(t)$ at
short times is qualitatively different. At long times, however, both functions
grow linearly with time. Comparing the insets in Figs. \ref{msd} and \ref{mobility}
we see that, although the persistence time dependence of $D$ and $\mu$ is 
qualitatively similar, there are significant quantitative differences. 

We recall that in the limit of the vanishing persistence time our system becomes
equivalent to a Brownian system at temperature $T=T_a$. Thus, we expect that
in the $\tau_p\to 0$ limit the ratio $D/\mu$ should be approaching $T_a$. 
For longer persistence times we \emph{define} an effective temperature based on the 
Einstein relation as
\begin{equation}\label{TeffE}
T_{\mathrm{eff}}^{\mathrm{E}} = D/\mu.
\end{equation}
The difference between the active temperature $T_a$ and the effective
temperature $T_{\mathrm{eff}}^{\mathrm{E}}$ 
based on the Einstein relation quantifies the departure of
our active matter system from equilibrium \cite{commentlowd}.

\begin{figure}
\includegraphics[width=1.0\columnwidth]{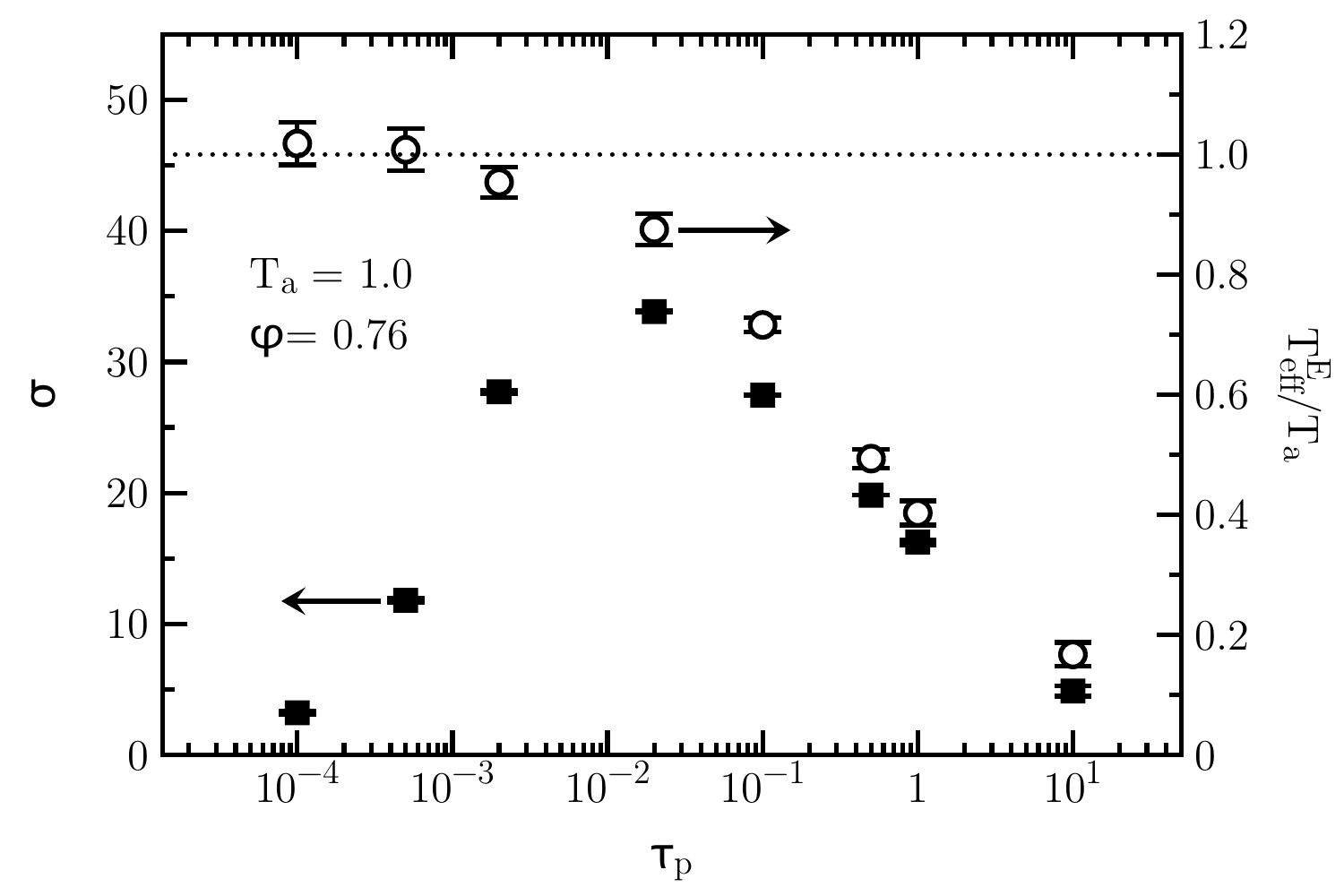}\\
\includegraphics[width=1.0\columnwidth]{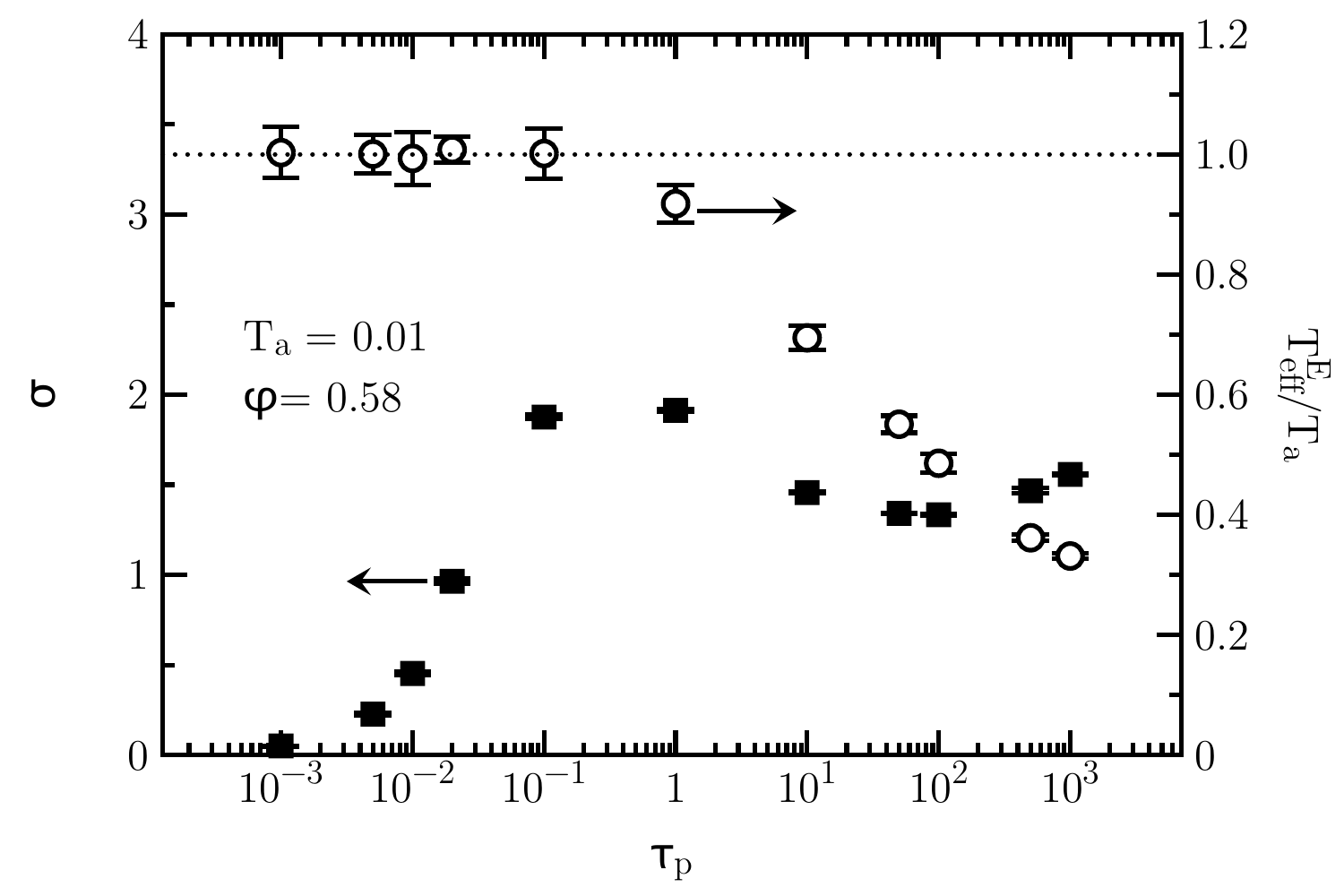}
\caption{\label{ratio}
The persistence time dependence of the ratio of the effective and 
active temperatures, $T_{\mathrm{eff}}^{\mathrm{E}}/T_a$ 
(open circles) and of the entropy production $\sigma$
as defined by Fodor \textit{et al.} \cite{Fodor2016} (filled squares). 
The ratio $T_{\mathrm{eff}}^{\mathrm{E}}/T_a$ monotonically
decreases with increasing persistence time whereas $\sigma$ exhibits a 
non-monotonic dependence on the persistence time. }
\end{figure}

In Fig. \ref{ratio} we show the persistence time dependence of the 
ratio of the effective temperature $T_{\mathrm{eff}}^{\mathrm{E}}$ based on 
the Einstein relation and the active temperature $T_a$.
We observe that for sufficiently short persistence times 
$T_{\mathrm{eff}}^{\mathrm{E}}/T_a$ is constant and equal to 1 (within error bars). 
With increasing persistence time this ratio starts decreasing monotonically 
\cite{commentL}. This behavior agrees with physical expectations and a
small $\tau_p$ expansion \cite{Fodor2016}.
For short persistence times the system is in an effective equilibrium state 
that can be described in terms of the active temperature, but with increasing 
persistence time the system is progressively displaced away from equilibrium
\cite{commentSL}. 

We note that the dependence of the ratio $T_{\mathrm{eff}}^{\mathrm{E}}/T_a$
on the persistence time is the opposite from the dependence on the shear rate 
of the ratio of the same effective temperature and the temperature $T$ for a 
sheared Brownian system. In the latter case, the ratio  
$T_{\mathrm{eff}}^{\mathrm{E}}/T$ increases monotonically \cite{Berthier2002,Szamel2011}
with increasing shear rate, \textit{i.e.} with increasing departure from equilibrium. 

\section{Entropy production rate} 
Fodor \textit{et al.} \cite{Fodor2016} started from
a definition of the entropy production in terms of a ratio of the probabilities
of a position space trajectory and its time-reversed version and derived the
following expression for the entropy production rate,
\begin{equation}\label{epr}
\sigma = \frac{\tau_p^2}{2 T_a} 
\left<\left(\sum_i \dot{\mathbf{r}}_i \cdot \partial_{\mathbf{r}_i}\right)^3 
\sum_{k>l} V(r_{kl})  \right>.
\end{equation}
We note that this expression involves third derivatives of the potential energy, and
in order to avoid any problems associated with singular contributions we
used an interparticle potential that has continuous first three derivatives.

We evaluated the persistence time dependence of the entropy production rate 
given by expression (\ref{epr}) for both state points. The results shown in Fig. 
\ref{ratio} do not follow our physical expectation. The quantity given by expression 
(\ref{epr}) has a non-monotonic dependence on the persistence time, and, moreover, this 
dependence is qualitatively different for the two active temperatures $T_a$. 
 
For both $T_a$, at small persistence times $\sigma$ increases with 
increasing persistence time, as predicted in Ref. \cite{Fodor2016}. 
At $T_a=1.0$, at the persistence time at which 
the ratio $T_{\mathrm{eff}}^{\mathrm{E}}/T_a$ is about 0.8, $\sigma$ exhibits
a maximum and then starts to decrease for longer persistence times.
At $T_a=0.01$, we observe initially a very similar persistence time dependence,
with the maximum occurring for the persistence time at which the ratio
$T_{\mathrm{eff}}^{\mathrm{E}}/T_a$ is about 0.9. However, in this case 
for longer persistence times $\sigma$ exhibits a local minimum and then it starts
to increase again. We note that the range of persistence times investigated
at $T_a=1.0$ is smaller than that investigated at $T_a=0.01$. The reason is that 
at $T_a=1.0$ at persistence times longer than $\tau_p=10$ very slow dynamics prevented
us from reaching a stationary state. 

We note that the relation between the location of the maximum of the entropy
production as a function of the persistence time and the value of the ratio
$T_{\mathrm{eff}}^{\mathrm{E}}/T_a$ is not necessarily universal and deserves further study. 

\section{Discussion} 
Our physical expectation is 
that increasing persistence time displaces an AOUP system progressively
away from equilibrium. This is supported by the monotonically increasing non-trivial
equal-time correlations between AOUPs velocities and the monotonically decreasing
ratio of the effective temperature based on the Einstein relation and the 
active temperature. Surprisingly, the dependence of the entropy production rate 
calculated according to Fodor \textit{et al.} on the persistence time does not
agree with this expectation. This suggests that expression (\ref{epr}) is 
not a good quantitative measure of the departure from equilibrium.

One may observe that the entropy production rate should be normalized by
a quantity characterizing system's dynamics, \textit{i.e.} that instead of $\sigma$
one should examine $\sigma\tau_\text{rel}$, where $\tau_\text{rel}$ is 
a characteristic relaxation time of the system. $\sigma\tau_\text{rel}$ 
quantifies the entropy produced while the system 
decorrelates from its given configuration. We used the characteristic diffusion 
time $\sigma_{BB}^2/D$, where $\sigma_{BB}$ is the $B$ particle size, 
as a measure of the relaxation time and we examined $\sigma\sigma_\text{BB}^2/D$
as a function of the persistence time. We found that $\sigma\sigma_\text{BB}^2/D$
monotonically increases with $\tau_p$ at $T_a=1.0$ but has a non-monotonic
dependence on $\tau_p$ at $T_a=0.01$. 

We recall that expression (\ref{epr}) also predicts vanishing entropy production
for a single freely moving AOUP and for a single AOUP in a harmonic potential.
Both findings are often claimed to be counter-intuitive. Some of the alternative
approaches to the entropy production 
\cite{Mandal2017,Dabelow2019,Shankar2018,Szamel2019}
find non-vanishing entropy production for a freely moving AOUP and/or
for an AOUP in a harmonic potential. It would be interesting to check what do these
approaches predict for systems of interacting AOUPs. 

Finally, we note that approaches of Refs. \cite{Dabelow2019,Shankar2018,Szamel2019}
either implicitly or explicitly consider trajectories in the space of
positions and self-propulsions. Thus, at least in spirit, they are similar to the 
approach of Pietzonka and Seifert \cite{Pietzonka2018} who argue that, 
in order to properly evaluate the entropy production, one has to consider
the physico-chemical processes giving rise to the self-propulsion. 
It would be very interesting to investigate whether
the minimal approach adopted in Refs. \cite{Dabelow2019,Shankar2018,Szamel2019} 
is sufficient to define and evaluate the entropy production 
or whether one has to follow the full treatment of 
Ref. \cite{Pietzonka2018}.

\section*{Acknowledgments}
We thank \'Etienne Fodor for comments on this work. 
We gratefully acknowledge the support of NSF Grants 
DMR-1608086 and CHE-1800282.


\begin{thebibliography}{99}

\bibitem{Bechingerrev} C. Bechinger, R. Di Leonardo, H. L\"owen, C. Reinchhardt,
G. Volpe and G. Volpe, ``Active particles in complex and crowded environments'',
Rev. Mod. Phys. \textbf{88}, 045006 (2016).

\bibitem{Marchettirev2} E. Fodor and M.C. Marchetti, 
``The statistical physics of active matter: From self-catalytic
colloids to living cells'', Physica A \textbf{504}, 106 (2018).

\bibitem{tenHagen} B. ten Hagen, S. van Teeffelen and H. L\"owen,
``Brownian motion of a self-propelled particle'', 
J. Phys.: Condens. Matter \textbf{23} 194119 (2011).

\bibitem{FilyMarchetti} Y. Fily and M.C. Marchetti,
``Athermal Phase Separation of Self-Propelled Particles with No Alignment'', 
Phys. Rev. Lett. \textbf{108}, 235702 (2012).

\bibitem{Szamel2014} G. Szamel, ``Self-propelled particle in an external potential: 
Existence of an effective temperature'', Phys. Rev E \textbf{90}, 012111 (2014).

\bibitem{Maggi2015} C. Maggi, U.M.B. Marconi, N. Gnan, and R. Di Leonardo,
``Multidimensional stationary probability distribution for interacting active 
particles'',
Scientific Reports \textbf{5}, 10742 (2015).

\bibitem{Fodor2016}
E. Fodor, C. Nardini, M. E. Cates, J. Tailleur, P. Visco, and F. van Wijland,
``How Far from Equilibrium Is Active Matter?'',
Phys. Rev. Lett. \textbf{117}, 038103 (2016).

\bibitem{Das2014} S.K. Das, S.A. Egorov, B. Trefz, P. Virnau, and K. Binder, 
``Phase Behavior of Active Swimmers in Depletants: Molecular Dynamics and
Integral Equation Theory'', 
Phys. Rev. Lett. \textbf{112}, 198301 (2014).

\bibitem{FarageKB} T.F.F. Farage, P. Krinninger, and J.M. Brader,
``Effective interactions in active Brownian suspensions'', 
Phys. Rev. E \textbf{91}, 042310 (2015).

\bibitem{noneqeq} For example, a recent report by Caprini \textit{et al.} 
\cite{Caprini2020} shows that motility-induced phase separation \cite{Marchettirev2} 
(MIPS) is accompanied by the appearance of long-range equal time velocity correlations, 
which are absent in equilibrium. According to the authors, this suggests that instead 
of a mapping of an active system with repulsive interaction onto a 
``passive'' thermal system with repulsive and attractive interactions,
a ``purely nonequilibrium approach'' is needed to describe MIPS.

\bibitem{Caprini2020} L. Caprini, U.M.B. Marconi, and A. Puglisi,
``Spontaneous Velocity Alignment in Motility-Induced Phase Separation'',
Phys. Rev. Lett. \textbf{124}, 078001 (2020).

\bibitem{Li2019} J. Li, J.M. Horowitz, T.R. Gingrich, and N. Fakhri,
``Quantifying dissipation using fluctuating currents'',
Nature Communications \textbf{10}, 1666 (2019).

\bibitem{Marconi2017} U.M.B. Marconi, A. Puglisi and C. Maggi,
``Heat, temperature and Clausius inequality in a model for active Brownian particles'',
Sci. Rep. \textbf{7}, 46496 (2017).

\bibitem{Caprini2019} L. Caprini, U.M.B. Marconi, A. Puglisi and A. Vulpiani,
``The entropy production of Ornstein-Uhlenbeck active particles: a path integral 
method for correlations'',
J. Stat. Mech. 053203 (2019).

\bibitem{Mandal2017} D. Mandal, K. Klymko and M.R. DeWeese,
``Entropy production and fluctuation theorems for active matter'',
Phys. Rev. Lett. \textbf{119}, 258001 (2017).

\bibitem{Dabelow2019} L. Dabelow, S. Bo and R. Eichhorn, 
``Irreversibility in Active Matter Systems: Fluctuation Theorem and Mutual Information'',
Phys. Rev. X \textbf{9}, 021009 (2019).

\bibitem{Shankar2018} S. Shankar and M.C. Marchetti,
``Hidden entropy production and work fluctuations in an ideal active gas'',
Phys. Rev. E \textbf{98}, 020604(R) (2018).

\bibitem{Szamel2019} G. Szamel, 
``Stochastic thermodynamics for self-propelled particles'',
Phys. Rev. E \textbf{100}, 050603(R) (2019).

\bibitem{Pietzonka2018} P. Pietzonka and U. Seifert, 
``Entropy production of active particles and for
particles in active baths'', 
J. Phys. A: Math. Theor. \textbf{51}, 01LT01 (2018).

\bibitem{BFS2017} L. Berthier, E. Flenner and G. Szamel,
``How active forces influence nonequilibrium glass transitions'',
New J. Phys. \textbf{19} 125006 (2017).

\bibitem{Garcia2015} S. Garcia, E. Hannezo, J. Elgeti, J.-F. Joanny, P. Silberzan, 
and N.S. Gov, ``Physics of active jamming during collective cellular
motion in a monolayer'', PNAS \textbf{112}, 15314 (2015).

\bibitem{SFB2015} G. Szamel, E. Flenner and L. Berthier, ``Glassy dynamics of athermal
self-propelled particles: Computer simulations and a nonequilibrium
microscopic theory'', Phys. Rev. E \textbf{91}, 062304 (2015).
 
\bibitem{Szamel2016} G. Szamel, 
``Theory for the dynamics of dense systems of athermal self-propelled
particles'', Phys. Rev. E \textbf{93}, 012603 (2016).

\bibitem{Cugliandolorev} L.F. Cugliandolo, ``The effective temperature'',
J. Phys. A: Math. Theor. \textbf{44}, 483001 (2011).

\bibitem{Szamel2017} G. Szamel, ``Evaluating linear response in active systems with no
perturbing field'', EPL \textbf{117}, 50010 (2017).

\bibitem{commentlowd} We note that in the small density (single-particle) limit
$T_{\mathrm{eff}}^{\mathrm{E}}=T_a$ for all persistence times. Thus, 
the ratio $T_{\mathrm{eff}}^{\mathrm{E}}/T_a$ is not a good indicator of the departure
from equilibrium of a free AOUP. 

\bibitem{commentL} We note that an earlier study \cite{LevisBerthier} reported
that the effective temperature based on the Einstein relation is increasing with
increasing persistence time of the persistent Monte Carlo dynamics. However,
to compare with our result one would have to investigate for the model
considered in Ref. \cite{LevisBerthier} the persistence time
dependence of the ratio of the effective temperature at finite density to 
that in the small density (single-particle) limit. Fig. 1c of Ref. \cite{LevisBerthier}
suggests that this ratio decreases with incresing persistence time. 

\bibitem{LevisBerthier} D. Levis and L. Berthier, 
``From single-particle to collective effective temperatures
in an active fluid of self-propelled particles'', 
EPL \textbf{111}, 60006 (2015).

\bibitem{commentSL} We also calculated self-diffusion, mobility coefficients
and effective temperatures for small and large particles separately. For the state 
points investigated these
individual effective temperatures are withing error bars of the effective temperatures
calculated using all the particles. 

\bibitem{Berthier2002} L. Berthier and J.-L. Barrat, 
``Shearing a Glassy Material: Numerical Tests of Nonequilibrium Mode-Coupling Approaches
and Experimental Proposals'', Phys. Rev. Lett. \textbf{89}, 095702 (2002); 
``Nonequilibrium dynamics and fluctuation-dissipation relation
in a sheared fluid'', J. Chem. Phys. \textbf{116}, 6228 (2002).

\bibitem{Szamel2011} G. Szamel and M. Zhang,
``Tagged particle in a sheared suspension: Effective temperature
determines density distribution in a slowly varying external potential
beyond linear response'', EPL \textbf{96}, 50007 (2011).


\end{thebibliography}
\end{document}